\documentclass[%
aip,amsmath,amssymb,
reprint,
author-numerical,
hidelinks]{revtex4-1}
\usepackage[utf8]{inputenc}
\usepackage[T1]{fontenc}
\usepackage{amsmath}
\usepackage{amsfonts}
\usepackage{amssymb}
\usepackage{empheq}
\usepackage{graphicx}
\usepackage{dcolumn}
\usepackage{mathptmx}
\usepackage{orcidlink}
\usepackage{hyperref}
\hypersetup{colorlinks, citecolor=cyan, linkcolor=cyan, urlcolor=cyan}

\usepackage{tikz}
\usetikzlibrary{shapes}
\usetikzlibrary{plotmarks}

\begin{document}

\title{Sensitivity of synchrotron radiation to the superthermal electron population in mildly relativistic plasma} 
\date{\today}
\author{M. E. Mlodik\orcidlink{0000-0003-4300-3941}} 
\email{mmlodik@princeton.edu}
\author{V. R. Munirov\orcidlink{0000-0001-6711-1272}}
\author{T. Rubin\orcidlink{0000-0001-6485-0096}}
\author{N. J. Fisch\orcidlink{0000-0002-0301-7380}}
\affiliation{Department of Astrophysical Sciences, Princeton University, Princeton, New Jersey, USA, 08540}
%and \\
%Princeton Plasma Physics Laboratory, Princeton, New Jersey, USA, 08540}

\begin{abstract}

Synchrotron radiation has markedly different behavior in $\sim 10~\textrm{keV}$ and in $\sim 100~\textrm{keV}$ plasma. We show that high-energy electrons which occupy the tail of velocity distribution function have disproportionate impact on power loss of $\sim 100~\textrm{keV}$ plasma. If electrons with energy more than a cutoff energy are redistributed while keeping the Maxwellian distribution function below cutoff energy intact, both emission and absorption of synchrotron radiation act to decrease the lost power. These novel radiation transport effects in non-equilibrium plasma suggest large utility in the deconfinement of high-energy electrons to reduce synchrotron radiation in applications where the radiation is deleterious.

\end{abstract}

\maketitle 

\section{Introduction}
\label{sec:intro}

Synchrotron radiation occurs whenever a charged particle moves in a magnetic field. Synchrotron emission and absorption in plasmas are of great importance both in fusion and astrophysical settings and thus have been studied extensively.~\cite{Trubnikov1959,Trubnikov1961,Bekefi1966, Bornatici1983, Bornatici1994, Albajar2007a,Albajar2007b,Arunasalam1992,Beard1961,Beard1962,Bordovitsyn1999,Denk2018,Denk2020,Drummond1961,Drummond1963,Fidone1989,Hirshfield1961,Krall1986,Lee2022,Lyubarskii1998,OConnor2005,Marcowith2003,Munirov2017a,Nassri1986,Nassri1988,Peratt2015,Robinson1985,Ginzburg1965, Ginzburg1969,Zheleznyakov1996}
The basic theory of synchrotron radiation can be found in Refs.~\onlinecite{Trubnikov1959,Trubnikov1961,Bekefi1966, Bornatici1983, Bornatici1994}. In particular, it was observed in the pioneering work of Trubnikov~\cite{Trubnikov1959} that self-absorption of synchrotron radiation by plasma is a crucial effect, resulting in more power radiated into higher harmonics. In fusion plasmas, electron cyclotron emission is used both to heat plasma~\cite{Jaeger1972, Lieberman1973, Ott1980, Bernstein1981, Erckmann1994, Lloyd1998, Wolf2019} and to diagnose it.~\cite{Fredrickson1986,Fisch1989, Janicki1993, Barrera2010, Liu2022} The synchrotron emission has also been proposed to maintain toroidal currents in tokamaks upon asymmetric reflection of the emission back into the tokamak~\cite{Dawson1982}, which then drives an rf current upon reabsorption in the plasma.~\cite{Fisch1987} In astrophysical plasmas synchrotron radiation is widespread~\cite{Ginzburg1965, Ginzburg1969,Zheleznyakov1996} and usually comes from nonthermal power-law energy distributions;~\cite{Condon2016} it plays a crucial role in the physics of pulsar magnetospheres,~\cite{Lyubarskii1998,OConnor2005} active galactic nuclei,~\cite{Biermann1987, Dermer1997,  Osmanov2010} supernova remnants,~\cite{Berezhko2004, Ballet2006} and dominates the radio emission from normal galaxies.~\cite{Lang1969,Condon1992}
Note, however, that the synchrotron radiation spectrum behavior changes significantly once plasma is mildly relativistic. 

In this paper, we show that the removal of superthermal electrons, accompanied by redistribution of these electrons as thermal electrons, has a large effect on the radiation transport. It is not the objective of this paper to inquire into how these superthermal electrons are selectively removed from the plasma; we assume that various devices may be be employed to render these electrons less well confined.  It is our objective to explore the sensitivity  on the net emission from the plasma of the absence of these electrons, given that these electrons play an outsized role on both emission and absorption.

The superthermal electrons both emit more radiation and emit disproportionally more radiation into higher harmonics. As many plasmas are optically thick for lower harmonics of synchrotron radiation and optically thin for higher harmonics, the redistribution of superthermal electrons dramatically decreases synchrotron power losses from plasma. Moreover, if we keep the total number of electrons constant, self-absorption of synchrotron radiation also changes once superthermal electrons are redistributed to lower energy. Note that this change of absorption enhances the effect of distribution function manipulation even more. 
This analysis is of potential importance for any device which aims to have magnetically confined high-temperature plasma where synchrotron losses are to be avoided or mitigated.

The paper is organized as follows. Formulation of the problem as well as governing equations are described in Sec.~\ref{sec:formulation}. The main results of the paper are shown in Sec.~\ref{sec:result}. Conclusions and limitations of the analysis are presented in Sec.~\ref{sec:discussion}.

%There are significant advantages of pursuing p-B11 fusion such as aneutronic character of reaction and abundance of fuel. However, the nature of cross-section of fusion reaction between proton and boron necessitates high temperatures ($T_e \gtrsim 100~$). At these temperatures, synchrotron radiation spectrum is markedly different from radiation of conventional burning D-T plasmas ($T_e \sim 10~$). Specifically, it is shown in Figure~\ref{fig:synchrotronSpectrum}. As such, mitigation of synchrotron radiation is necessary for a hot fusion reactor to reach ignition.

\section{Problem formulation}
\label{sec:formulation}

In order to isolate the effect of superthermal electrons on power losses of plasma via synchrotron radiation, we consider the following model. Suppose that there is a uniform plasma slab immersed in uniform magnetic field $B$ parallel to the boundary of the slab. Suppose that the electron distribution function in this slab is the same everywhere and that it is given. For the most part, unless said otherwise, we consider perpendicular propagation of synchrotron radiation as most losses are concentrated around that direction of propagation. We also consider the plasma slab to be tenuous to synchrotron radiation (this assumption is discussed in more detail in Sec.~\ref{sec:discussion}).

% Given that the collision frequency in plasma decreases with increase in relative velocity, non-Maxwellian features of electron distribution function are easier to sustain in mildly relativistic plasma than in non-relativistic plasma.

More specifically, we consider a relativistic Maxwellian distribution with and without a cutoff energy. We will use the Maxwell--Jüttner distribution with
a cutoff at $\gamma_{max}$: 
\begin{equation}
f\left(\mathbf{u}\right)=\begin{cases}
N_{const}\frac{e^{-\frac{\gamma}{\theta_{T_{e}}}}}{4\pi\theta_{T_{e}} K_{2}\left(1/\theta_{T_{e}}\right)}, & \gamma\leq\gamma_{max},\\
0, & \gamma>\gamma_{max},
\end{cases}\label{eq:MJ}
\end{equation}
where $N_{const}$ is the normalization constant, $K_{2}$ is the modified Bessel function of the second kind, $\theta_{T_{e}}=T_{e}/m_{e}c^{2}$ is the dimensionless temperature, and $\gamma$ is the Lorentz factor. Throughout this paper we will use the dimensionless momentum $\mathbf{u}=\mathbf{p}/m_{e}c$, so that $\gamma=\sqrt{1+u^{2}}$; the
normalization will always be such that $\int f\left(\mathbf{u}\right)d\mathbf{u}=\int f\left(\mathbf{u}\right)2\pi u_{\perp}du_{\perp}du_{\parallel}=1$, and by subscripts $\perp$ and $\parallel$ we will denote the components perpendicular and parallel to the magnetic field. When $\gamma_{max}$ is equal to infinity and $N_{const}=1$, Eq.~(\ref{eq:MJ}) yields the Maxwell--Jüttner distribution without cutoff.
This family of model distributions is chosen to show with the most clarity the impact of superthermal electrons on synchrotron radiation.

Synchrotron radiation in tenuous plasma can be described as follows. If the spontaneous emissivity from one electron with momentum $\mathbf{u}$ is given by $\eta_{\omega}\left(\mathbf{u}\right)$,
then the total emission coefficient from a collection of electrons of density $n_{e}$ with distribution
function $f\left(\mathbf{u}\right)$ is given by~\cite{Bekefi1966}
\begin{multline}
j_{\omega}=n_{e}\int\eta_{\omega}\left(\mathbf{u}\right)f\left(\mathbf{u}\right)d\mathbf{u}\\
=n_{e}\int\eta_{\omega}\left(u_{\perp},u_{\parallel}\right)f\left(u_{\perp},u_{\parallel}\right)2\pi u_{\perp} du_{\perp}du_{\parallel}.\label{eq:emission_j}
\end{multline}

The absorption coefficient is given by~\cite{Bekefi1966}
\begin{multline}
\alpha_{\omega}=-n_{e}\frac{8\pi^{3}c^{2}}{n_{r}^{2}\omega^{2}}\int\eta_{\omega}\left(u_{\perp},u_{\parallel}\right)\left[\frac{\varepsilon}{m_{e}^{2}c^{4} }\frac{\partial f}{\partial u_{\perp}}\right.\\
\left.-\frac{n\left(\theta\right)\cos\theta}{m_{e}c^{2}}\left(u_{\parallel}\frac{\partial f}{\partial u_{\perp}}-u_{\perp}\frac{\partial f}{\partial u_{\parallel}}\right)\right]2\pi du_{\perp}du_{\parallel}\\
=-n_{e}\frac{8\pi^{3}c^{2}}{n_{r}^{2}\omega^{2}}\int\eta_{\omega}\left(\mathbf{u}\right)\frac{\partial f\left(\mathbf{u}\right)}{\partial\varepsilon}d\mathbf{u}.\label{eq:absorption_alpha}
\end{multline}
\noindent Here, $\theta$ is the angle between the wave propagation and the magnetic field, $n_{r}$ is the ray-refractive index [see Eq.~(1.121)
of Ref.~\onlinecite{Bekefi1966}], while $n$ is the usual wave refractive index. We will use the assumption of tenuous plasma and set $n_{r}=n=1$.

The emissivity of a single electron from tenuous plasma is given by~\cite{Trubnikov1961}

\begin{multline}
\eta_{\omega}=\frac{e^{2}\omega^{2}}{2\pi c\omega_{c}\sin^{2}\theta\gamma}\sum_{s=1}^{\infty}\left[\left(\gamma\cos\theta-u_{\parallel}\right)^{2}J_{s}^{2}\left(\frac{\omega\sin\theta}{\omega_{c}}u_{\perp}\right)\right.\\
\left.+u_{\perp}^{2}\sin^{2}\theta J_{s}^{\prime2}\left(\frac{\omega\sin\theta}{\omega_{c}}u_{\perp}\right)\right]\\
\times\delta\left[s-\frac{\omega}{\omega_{c}}\left(\gamma-u_{\parallel}\cos\theta\right)\right].\label{eq:eta_emissivity}
\end{multline}

Here, $\omega_c= |e|B/(m_e c)$ is the electron cyclotron frequency in the non-relativistic limit. This formula is separated into two parts (the first is proportional
to $J_{s}^{2}$ and the second is proportional to $J_{s}^{\prime2}$)
that correspond to two separate polarizations. The corresponding total
plasma emission and absorption determined by Eqs.~(\ref{eq:emission_j})\textendash (\ref{eq:absorption_alpha})
with $\eta_{\omega}$ given by Eq.~(\ref{eq:eta_emissivity}) are
also separated into two polarizations, which we will denote with superscripts (1) and (2). Separation above might not be possible for absorption, as can be seen, for example, by looking at the $\left|\mathbf{e}\cdot\mathbf{V}_{n}^{*}\right|^{2}$
term in Refs.~\onlinecite{Bornatici1994, Albajar2007a}.
However, this separation is legitimate for near perpendicular propagation
($\theta\approx\pi/2$), which dominates radiation losses
for such temperatures. Moreover, the polarization should affect the
distribution functions with cutoff and without cutoff in a similar
manner (since both of them are symmetric). In the case of perpendicular propagation, superscripts (1) and (2) correspond to ordinary and extraordinary wave, respectively.

For the radiation intensity leaving a slab of plasma per $d\omega$ and per solid angle, the equation of radiative transfer yields the following expression  {[}see, for example, Eq.~(10) of Ref.~\onlinecite{Trubnikov1959}{]}:

\begin{equation}
I_{\omega}=\sin\theta\left[\frac{j_{\omega}^{\left(1\right)}}{\alpha_{\omega}^{\left(1\right)}}\left(1-e^{-\frac{\alpha_{\omega}^{\left(1\right)}L}{\sin\theta}}\right)+\frac{j_{\omega}^{\left(2\right)}}{\alpha_{\omega}^{\left(2\right)}}\left(1-e^{-\frac{\alpha_{\omega}^{\left(2\right)}L}{\sin\theta}}\right)\right],\label{eq:I_w}
\end{equation}
\noindent where superscripts $(1)$ and $(2)$ correspond to two polarizations
discussed above. Note that the crucial parameters that determine whether plasma is opaque for a given frequency is $\alpha_{\omega}^{\left(1,2\right)} L$. Following Ref.~\onlinecite{Bekefi1966}, in plasma with a symmetric distribution function that depends only
energy, $\alpha_{\omega}^{\left(1,2\right)} L$ have the following dependence:
\begin{gather}
    \alpha_{\omega}^{\left(1,2\right)} L = \frac{\omega_p^2}{\omega_c} \frac{L}{c} \sum_{s=1}^{\infty} \Phi^{\left(1,2\right)} \left(s; \omega/\omega_c; {\theta_{T_{e}}} \right),
\end{gather}
where $\Phi^{\left(1,2\right)}$ are dimensionless functions. As $\Phi^{\left(1,2\right)}$ do not depend on plasma size or density, plasma absorption is determined to a large extent by absorption parameter $\Lambda$:
\begin{gather}
    \Lambda = \frac{\omega_p^2}{\omega_c} \frac{L}{c}.
    \label{eq:Lambda}
\end{gather}
Here, $\omega_p$ is plasma frequency, $L$ is plasma size in the direction perpendicular to the magnetic field.
Note that the redistribution of superthermal electrons which is studied in this paper does not change $\Lambda$.

For the Maxwell\textendash Jüttner distribution with a cutoff energy,
we still have a symmetric distribution function that depends only
energy, so that $\partial f\left(\mathbf{u}\right)/\partial\varepsilon=-f\left(\mathbf{u}\right)/\theta_{T_{e}}m_{e}c^{2}$,
and the source function is the same as for the blackbody radiation:
\begin{equation}
\frac{j_{\omega}}{\alpha_{\omega}}=\frac{j_{\omega}^{\left(1\right)}}{\alpha_{\omega}^{\left(1\right)}}=\frac{j_{\omega}^{\left(2\right)}}{\alpha_{\omega}^{\left(2\right)}}=\frac{\omega^{2}T_{e}}{8\pi^{3}c^{2}}=\frac{m_{e}\omega^{2}\theta_{T_{e}}}{8\pi^{3}}.\label{eq:source_function}
\end{equation}

Using source functions from Eq.~(\ref{eq:source_function}), we can
rewrite Eq.~(\ref{eq:I_w}) and obtain (see also Section 11.9 of
Ref.~\onlinecite{Krall1986}):
\begin{equation}
I_{\omega}=\sin\theta\frac{m_{e}\omega^{2}\theta_{T_{e}}}{8\pi^{3}}\left(2-e^{-\frac{\alpha_{\omega}^{\left(1\right)}L}{\sin\theta}}-e^{-\frac{\alpha_{\omega}^{\left(2\right)}L}{\sin\theta}}\right).
\end{equation}
The limiting case $\alpha_{\omega}^{\left(1,2\right)}L/\sin\theta\ll1$ of Eq.~(\ref{eq:I_w}) corresponds to the optically thin regime:
\begin{equation}
I_{\omega}=\left(j_{\omega}^{\left(1\right)}+j_{\omega}^{\left(2\right)}\right)L=j_{\omega}L.\label{eq:Iw_thin}
\end{equation}

The opposite case $\alpha_{\omega}^{\left(1,2\right)}L/\sin\theta\gg1$
corresponds to the optically thick regime, where radiation intensity becomes
\begin{equation}
I_{\omega}=\sin\theta\frac{m_{e}\omega^{2}\theta_{T_{e}}}{4\pi^{3}},
\end{equation}
i.e. blackbody spectrum. Note that $\alpha_{\omega}^{\left(1,2\right)}$ vary across the spectrum. Thus, the plasma can be optically thick for some frequencies and optically thin for other frequencies.

\section{Main result}
\label{sec:result}

%While net power loss via synchrotron radiation can be found using Eqs.~(\ref{eq:emission_j})\textendash (\ref{eq:absorption_alpha}) with the individual emissivity given by Eq.~(\ref{eq:eta_emissivity}), the order of integration matters. %One strategy is to get rid of the delta function by first integrating over $du_{\parallel}$ and then integrate the result over $du_{\perp}$ (or $d\gamma$).
%The problem with this approach is that it does not work for $\theta$ exactly $\pi/2$, since we have to divide by $\cos\theta$ when we integrate delta function over $du_{\parallel}$. 
%The second approach is to get rid of the delta function by first integrating over $du_{\perp}$; this approach is described in Ref.~\onlinecite{Albajar2007a}.
%We implemented both approaches and they agree in the area of their applicability.

\begin{figure}[h]
	\includegraphics[width=0.48\textwidth]{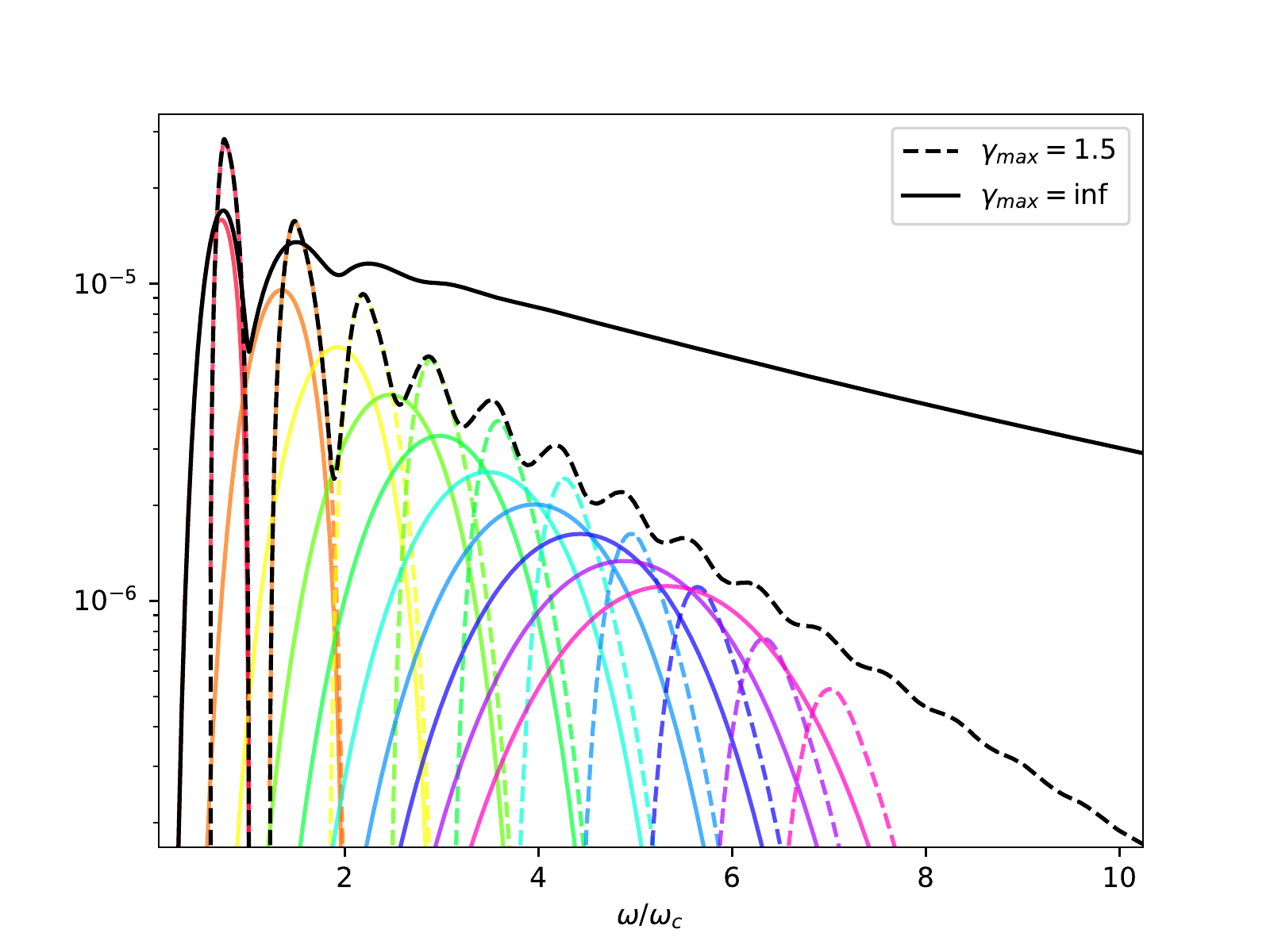}
	\centering	
	\caption{The total plasma emission coefficient $j_{\omega}$ for $\theta=0.45\pi$ calculated for 300 harmonics as a function of $\omega/\omega_{c}$ for $\gamma_{max}=1.5$ (black dashed line) and without cutoff (black solid line) together with their first 10 harmonics (color solid and dashed lines).}
	\label{fig:synchrotronEmission}	
\end{figure}

Both emission and absorption change significantly in a plasma with energy cutoff. An example of the emission spectrum determined by Eq.~(\ref{eq:emission_j}) for near perpendicular propagation angle $\theta=0.45\pi$ is shown in Fig.~\ref{fig:synchrotronEmission}. Once electrons with energy higher than cutoff energy are redistributed into low-energy part of the electron distribution function, emission into the low-frequency part of the spectrum increases while emission into high-frequency part of the spectrum decreases. This can be seen from the comparison of the total plasma emission coefficient of the Maxwell--Jüttner distribution without cutoff (black solid line) and the Maxwell--Jüttner distribution with cutoff at $\gamma_{max} = 1.5$ (black dashed line). 
Note that many ($300$) harmonics were used in order to calculate total emission coefficient because of line broadening, which is a large effect in mildly relativistic plasma due to both relativistic change of mass and Doppler broadening. As line broadening is less pronounced in plasma with energy cutoff, the larger part of the total emission coefficient spectrum retains oscillatory nature in such a plasma. Everywhere in this paper, a sufficient number of harmonics is used in calculations in order to properly capture all the relevant details of the spectrum.

\begin{figure}[h]
	\includegraphics[width=0.48\textwidth]{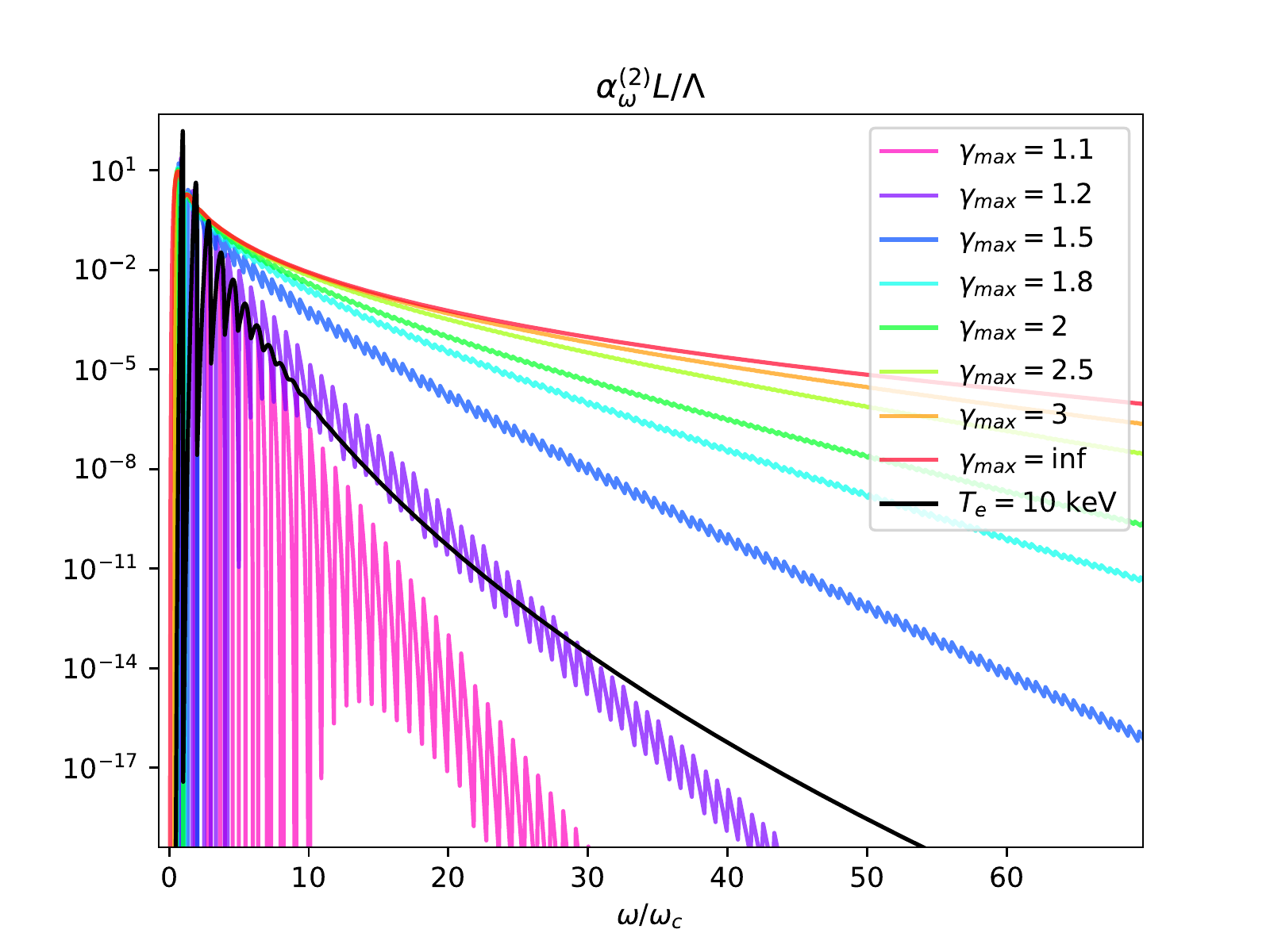}
	\centering	
	\caption{Synchrotron radiation absorption as a function of harmonic number $\omega/\omega_c$ for $10~\textrm{keV}$ plasma (black line) and $150~\textrm{keV}$ plasma for different values of the energy cutoff $\gamma_{max}$ (color lines).}
	\label{fig:synchrotronAbsorption}	
\end{figure}

The absorption spectrum (more specifically, spectrum of $\alpha_{\omega}^{(2)} L/\Lambda$) is shown in Fig.~\ref{fig:synchrotronAbsorption}. This particular combination of parameters is chosen in order to make the plot independent of plasma density and size. Color lines show absorption of plasma with temperature $150$ keV and energy cutoff at $\gamma = \gamma_{max}$ for different values of $\gamma_{max}$. Similarly to the emission spectrum, the absorption spectrum of plasma with energy cutoff features an increase at low frequencies and a dramatic decrease at high frequencies. 
Moreover, oscillations become more pronounced with decreasing cutoff energy. If all electrons with $\gamma > 1.2$ in $150$ keV plasma are redistributed into lower-energy part of the distribution function, the absorption spectrum becomes similar in magnitude to the spectrum of Maxwellian plasma at temperature $10$ keV (black line in Fig.~\ref{fig:synchrotronAbsorption}).
Absorption spectrum shows decrease with frequency up to oscillations regardless of the cutoff energy. As such, plasma remains more opaque to the low-frequency part of synchrotron radiation spectrum than to the high-frequency part.

\begin{figure}[h]
	\includegraphics[width=0.48\textwidth]{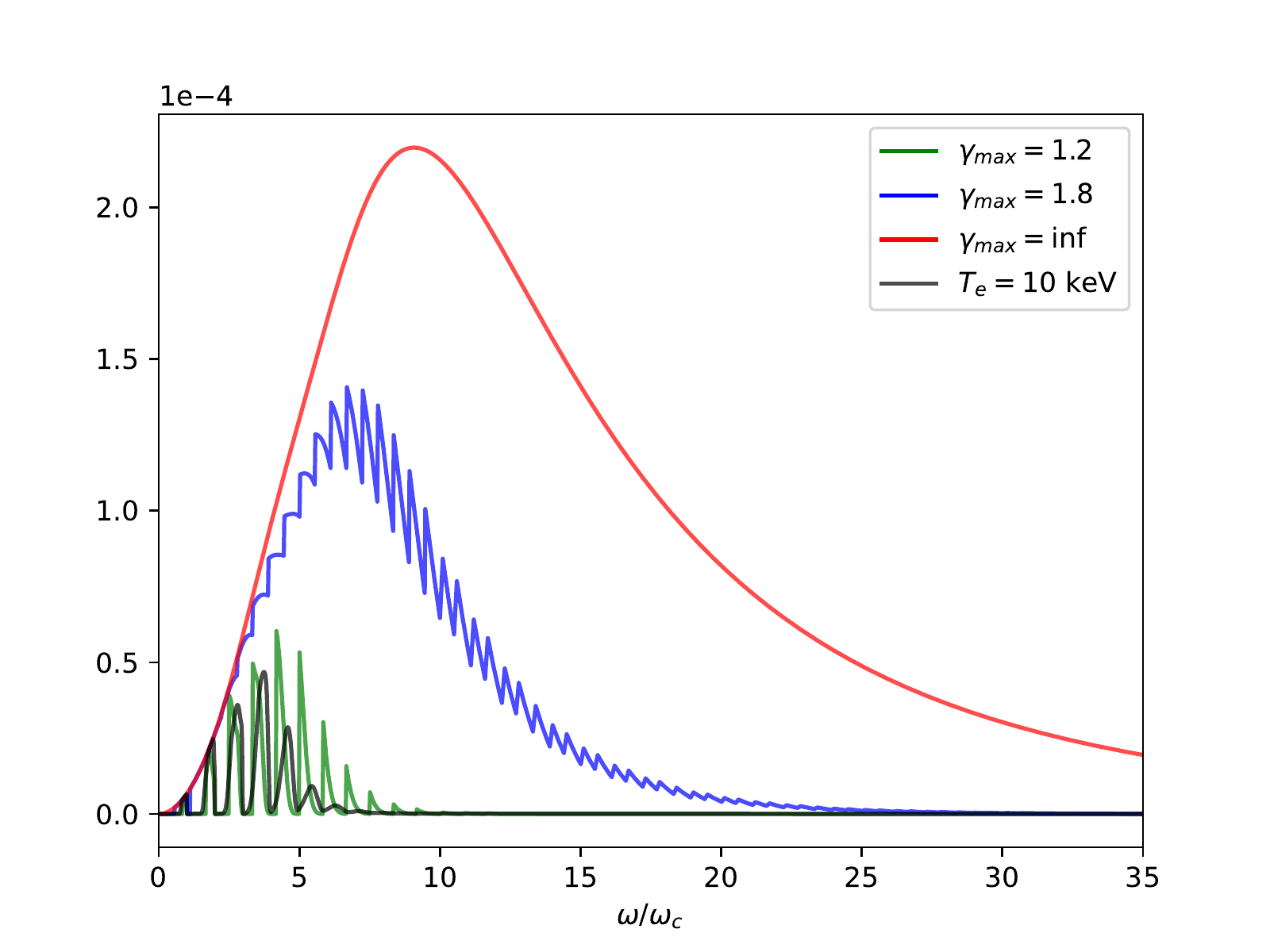}
	\centering	
	\caption{Synchrotron radiation spectrum in $10~\textrm{keV}$ plasma (black line), $150~\textrm{keV}$ plasma (red line), $150~\textrm{keV}$ plasma with electron energy cutoff at $\gamma_{max} = 1.8$ (blue line) and $\gamma_{max} = 1.2$ (green line). Here, opacity parameter $\Lambda = 600$ is assumed.}
	\label{fig:synchrotronSpectrum}	
\end{figure}

Combined, the changes in emission and absorption lead to the main observation described in this paper: superthermal electrons provide disproportionately large contribution to synchrotron energy losses in mildly relativistic plasma.  
One reason for this is that superthermal electrons radiate more energy than do bulk electrons and have worse single-particle energy confinement time. 
Another, even more important reason why superthermal electrons disproportionally affect synchrotron radiation is that they radiate more into higher harmonics. The nature of synchrotron radiation in hot plasma is such that low harmonics are primarily absorbed by the plasma itself (as long as $\Lambda \gg 1$) while high harmonics are radiated away. Let us write $\Lambda = 602~ n_{e,14} L_{1\textrm{m}} / B_{10\textrm{T}}$, where $n_{e,14}$ is electron density normalized to $10^{14}~\textrm{cm}^{-3}$, $L_{1\textrm{m}}$ is size of plasma in the direction perpendicular to the magnetic field normalized to $1~\textrm{m}$, and $B_{10\textrm{T}}$ is magnetic field normalized to $10~\textrm{T}$. Thus, if the plasma under consideration has perpendicular size $L = 1~\textrm{m}$, magnetic field $B = 10~\textrm{T}$, density $n_e = 10^{14}~\textrm{cm}^{-3}$, and temperature $T_e = 150~\textrm{keV}$, then $\Lambda = 602$. Therefore, if superthermal electrons are redistributed into the lower-energy part of the distribution function, then electrons radiate more into low-frequency part of the spectrum which is mostly absorbed by the plasma itself and they radiate less into high-frequency part of the spectrum, to which plasma is mostly transparent.
An example which corroborates this observation is shown in Fig.~\ref{fig:synchrotronSpectrum}. In Figure~\ref{fig:synchrotronSpectrum}, red line shows power loss spectrum, which includes both emission and self-absorption of $150$ keV plasma; blue line shows power loss spectrum of $150$ keV plasma with cutoff $\gamma_{max} = 1.8$; green line shows power loss spectrum of $150$ keV plasma with cutoff $\gamma_{max} = 1.2$; black line shows power loss spectrum of $10$ keV Maxwellian plasma. In all cases in Fig.~\ref{fig:synchrotronSpectrum}, $\Lambda = 600$ is assumed. As cutoff energy is decreased, spectrum peak moves to lower frequency, oscillations become more pronounced, and overall power loss, determined by area under the curve, decreases dramatically.

\begin{figure}[h]
	\includegraphics[width=0.48\textwidth]{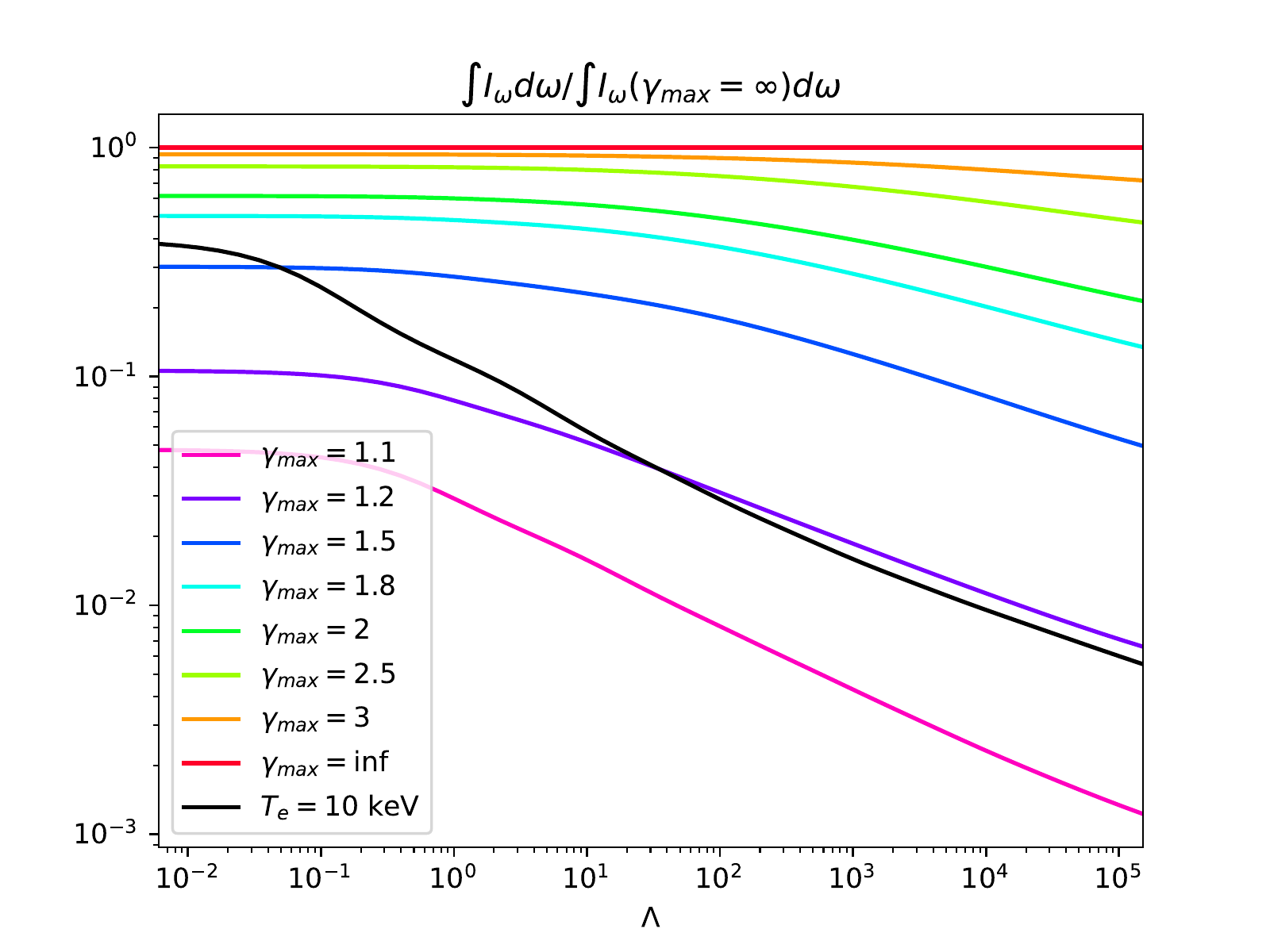}
	\centering	
	\caption{Synchrotron radiation mitigation via redistribution of superthermal electrons. Color lines show the ratio of total power loss via synchrotron radiation per electron as a function of opacity parameter $\Lambda$ for different values of cutoff parameter $\gamma_{max}$ in $150$~keV plasma. Black line shows the ratio of synchrotron power loss per electron in $10$~keV plasma to same quantity in $150$~keV, also as a function of opacity.}
	\label{fig:synchrotronMitigation}	
\end{figure}

Given all of the changes described above, the redistribution of superthermal electrons into lower-energy part of distribution function makes a disproportionate impact on power loss via synchrotron radiation. 
Figure~\ref{fig:synchrotronMitigation} shows the dependence of the ratio of synchrotron radiation power loss for electron distribution function with and without a cutoff on absorption parameter $\Lambda$. Note that the electron density is kept the same in Fig.~\ref{fig:synchrotronMitigation} so the figure shows the power loss per particle.
Limit $\Lambda \ll 1$ shows the effect of decrease in emission due to redistribution of superthermal electrons, as plasma is transparent to all harmonics in this limit. Power loss per particle decreases dramatically as cutoff energy decreases. Even more striking feature is that for $\Lambda < 10^5$, power loss decrease per electron becomes larger as absorption parameter $\Lambda$ decreases. In other words, redistribution of superthermal electrons is more effective for mitigation of power loss via synchrotron radiation in opaque plasma than in transparent plasma. For example, synchrotron power loss mitigation in plasma with cutoff at $\gamma_{max} = 1.5$ and $\Lambda \ll 1$ is as effective as mitigation in plasma with cutoff at $\gamma_{max} = 1.8$ and $\Lambda = 800$. Note that in order to decrease synchrotron power loss per electron by a factor of $2$, cutoff at $\gamma_{max} = 1.8$ is required in $\Lambda \ll 1$ plasma, cutoff at $\gamma_{max} = 2$ is sufficient in $\Lambda =60$ plasma, and cutoff at $\gamma_{max} = 2.5$ is enough in $\Lambda = 4 \times 10^4$ plasma if plasma temperature is $150$~keV. Another feature of note in Fig.~\ref{fig:synchrotronMitigation} is that power loss due to synchrotron radiation is a relatively larger issue in $150$ keV plasma than in $10$ keV plasma if opacity parameter is large, as shown by black curve. Therefore, even though the redistribution of superthermal electrons is more effective as opacity is increased, a larger redistribution (provided by smaller cutoff parameter $\gamma_{max}$) is required for $150$~keV plasma in order to match power loss of a $10$~keV plasma. As the cutoff energy goes to infinity, the redistribution of superthermal electrons becomes less and less effective as a method to decrease power loss due to synchrotron radiation, as fewer electrons get redistributed.

%If $\gamma_{max} = 2$ is taken as energy cutoff (i.e. electrons with energy higher than $511~\textrm{keV}$ are removed and same number of electrons with energy lower than cutoff energy is added), then in the limit $\Lambda \ll 1$ energy loss per electron is reduced by factor of $\sim 1.6$. 
%Given that in fusion device it is more realistic to expect $\Lambda = 10^2 ...10^4$, absorption provides additional benefit to the mitigation strategy proposed here. Moreover, should it be possible to move energy cutoff line to lower energies, the benefit of the strategy is going to increase even more. 
%The effect of phase space engineering of electron velocity distribution function can be understood by comparing synchrotron energy losses from two model distributions: Maxwell-Juttner (relativistic Maxwellian) distribution, which corresponds to basic scenario, and Maxwell-Juttner distribution with a cutoff at electron kinetic energy $\epsilon_{kin} = m_e c^2 \left( \gamma_{max} -1 \right)$, which corresponds to scenario with phase space engineering, with both distributions normalized to the same electron density.  

\begin{figure}[h]
	\includegraphics[width=0.48\textwidth]{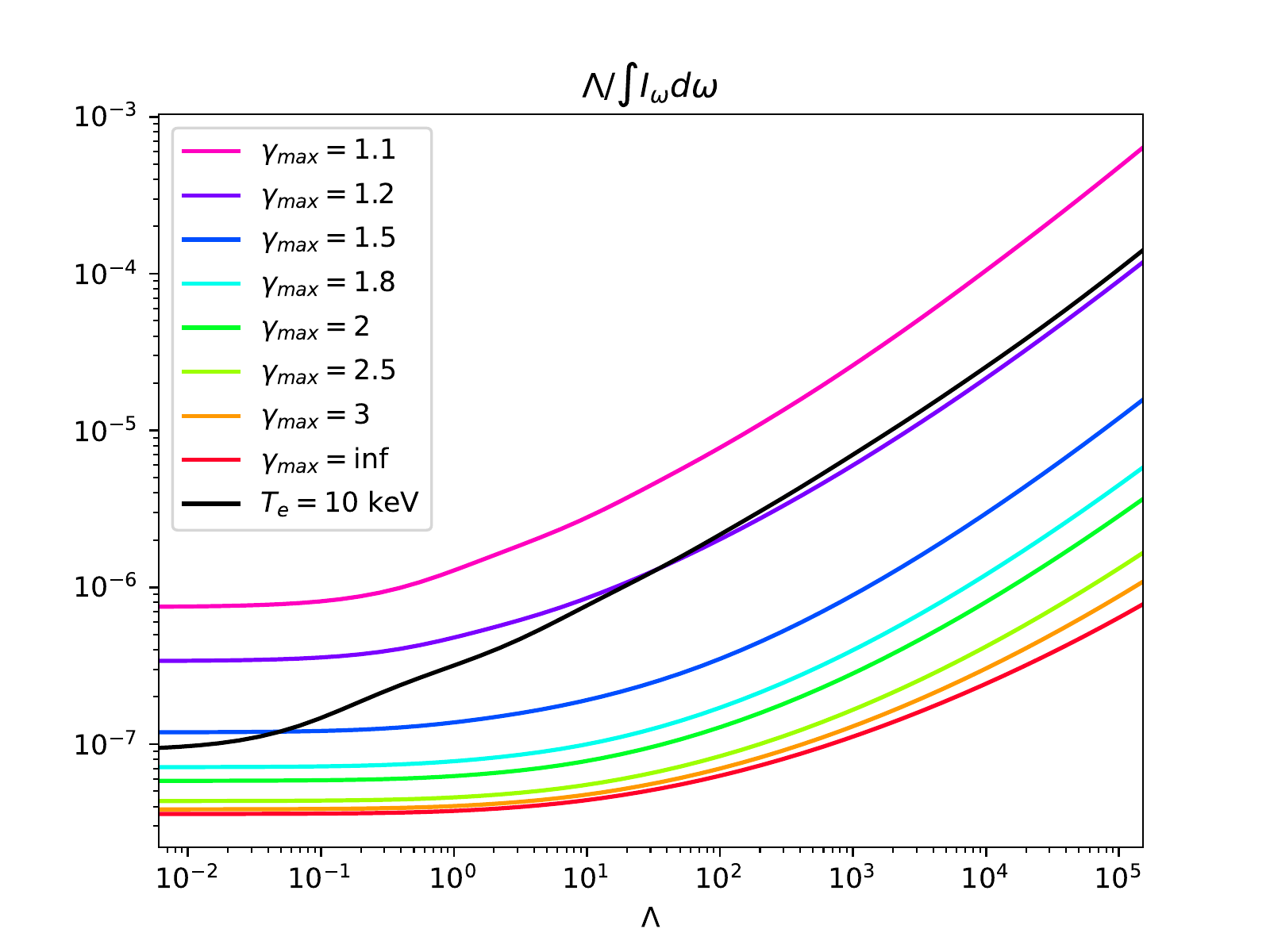}
	\centering	
	\caption{Synchrotron energy confinement time as a function of absorption coefficient $\Lambda$ and cutoff energy $\gamma_{max}$ in $150$~keV plasma (color lines) and energy confinement time in $10$~keV plasma (black line). Units are arbitrary.}
	\label{fig:synchrotronConfinementTime}	
\end{figure}

Another way to interpret power loss mitigation is to look at synchrotron energy confinement time, which can be defined as the ratio of kinetic energy stored in electrons to power loss via synchrotron radiation. Given that surface density of kinetic energy stored in electrons in a plasma slab is $n_e T_e L$, energy confinement time is proportional to $\Lambda / \int I_{\omega} d\omega$. The total improvement in synchrotron energy confinement time due to dependence on absorption coefficient and energy cutoff is shown in Figure~\ref{fig:synchrotronConfinementTime}.
The synchrotron energy confinement time increases both with increase in opacity parameter $\Lambda$ and with decrease in cutoff $\gamma_{max}$.

\section{Discussion and Caveats}
\label{sec:discussion}

Superthermal electrons have disproportional impact on all aspects of synchrotron radiation in plasma: emission and absorption spectra, power loss, and energy confinement time. 
The analysis presented in this paper is focused on a Maxwell\textendash Jüttner distribution with energy cutoff and the losses in the direction perpendicular to slab boundaries (which is the primary direction of power losses); it is chosen to isolate and cleanly demonstrate the effect. Similar effects can be expected for other electron distribution functions and other angles of propagation. The size of power loss reduction can be significant; for example, the power loss is halved in $150$ keV plasma with opacity parameter $\Lambda = 60$ and cutoff at $\gamma_{max} = 2$.

Reducing the power loss by synchrotron radiation in plasma with a $150$ keV electron temperature may be relevant for pB11 fusion concepts. For pB11 fusion to be viable, the plasma must be kept at a high temperature, where Bremsstrahlung losses could be overwhelming.  
It was recently shown that, neglecting synchrotron radiation and other conductive losses, the pB11 reaction could in principle achieve ignition by overcoming the Bremsstrahlung losses, even if barely, at electron temperatures in the range of $100-200$ keV.\cite{Putvinski2019} 
If some form of alpha channeling\cite{Fisch1992} were employed, this margin could be significantly widened.\cite{Kolmes2022,Ochs2022} 
With alpha channeling, the margin is most wide at electron temperatures of about $150$ keV. 
These calculations do not assume any particular means of plasma confinement; rather the focus is just to see to what extent the Bremsstrahlung losses could be overcome.
However, if the plasma were confined magnetically, where synchrotron losses could also be large, then the means provided here to reduce those losses in the $150$ keV electron temperature range could be relevant.

The results here show that deconfining superthermal electrons by any means significantly  reduces synchrotron emission.  It should be noted that the removal of these electrons in of itself is a significant energy loss.  These losses occur on a continuing basis, since  thermal electrons are  constantly promoted via collisions to superthermal elecrons.  Thus, the plasma energy losses are not stemmed by the deconfinement of superthermal electrons, only that the energy loss through synchrotron radiation is now lost instead through energetic particles.  However, in cases in which the radiation itself is damaging, or the energy lost in particles is more easily recovered than the energy lost in radiation, this effect can be beneficial.

These results are of particular importance for plasma devices with open-field-line geometry. 
In open-field-line plasma devices, there are several mechanisms to deplete the superthermal tail of the electron distribution function, for example scattering by imperfections of the magnetic field, or by magnetic turbulence.\cite{Rechester1978} This paper shows the utility of depletion of electron tail should it be realized, thereby providing an argument for particular design choices. Note that superthermal electrons are less collisional than bulk electrons, which, depending on the details of mechanism of the redistribution of the superthermal electrons, could lead to anisotropy in the electron distribution function. However, given the outsized and synergistic role of superthermal electrons in emission and absorption, it can be expected that the dramatic decrease of power loss via synchrotron radiation in the perpendicular direction found here will be retained.  Although anisotropy in the electron distribution function could affect the off-perpendicular propagation of synchrotron radiation, those details are beyond the scope attempted here.

While Figure~\ref{fig:synchrotronMitigation} captures the basic effect, more precise calculations might refine the results presented here by perhaps $\sim 10\%$. For example, we assumed that plasma is tenuous, which is satisfied if $\left(\omega_p/\omega_c\right)^2 \ll \theta_{T_e}$. If this condition is not satisfied, the low harmonics might not be able to propagate in plasma. Note, however, that taking this effect properly into account is only going to strengthen the claim described here. The reason is that while propagation of lower harmonics might be affected if plasma is not tenuous, higher harmonics are going to propagate as is. Given that the redistribution of superthermal electrons is predominantly affecting higher harmonics, the relative ratio of power losses is going to increase if lower harmonics cannot propagate in plasma. 
We also assumed that the plasma is homogeneous, while plasmas in nature or in laboratory settings are often inhomogeneous. Spatial inhomogeneity of plasma may give rise to effects such as formation of regions of plasma where electron tails are naturally formed,\cite{Harvey1981,Giruzzi1992} which can affect synchrotron losses from plasma. That could modify the magnitude of the effect reported on here but not the nature of the result.

% Phase space engineering required to properly exploit the invention disclosed here is indeed possible. For example, if one considers a centrifugal mirror machine with negative mirror ratio, where ions are confined centrifugally and electrons are confined electrostatically, low-energy electrons are confined preferentially as their kinetic energy is smaller than electrostatic potential energy. Conversely, high-energy electrons are not going to be confined by electrostatic potential. Moreover, if mirror ratio is negative, magnetic field is going to deconfine such electrons. Therefore, high-energy electrons are going to be promptly lost in such a device. Their population is going to be replenished via collisions, but it is a slow process because collision frequency is low for such a high energy electrons. The example above is a proof that phase space engineering of hot fusion plasma is possible but usefulness of the invention described in this patent disclosure is not limited to the example configuration.

\acknowledgments{The authors thank E. J. Kolmes, I. E. Ochs, and J.-M. Rax for useful conversations. This work was supported by ARPA-E Grant DE-AR0001554.}

\section*{Data Availability Statement}

Data sharing is not applicable to this article as no new data were created or analyzed in this study.

%\appendix
%\section{Extra calculations.}

%[Some calculations we might want to offload from main body of the paper.]


%\bibliographystyle{apsrev4-1} 
%\bibliography{synchrotron_losses}

\end{document}